**Clinical Courses of Acute Kidney Injury in Hospitalized Patients: A Multistate Analysis**

Esra Adiyeke, PhD[a,b], Yuanfang Ren, PhD[a,b], Ziyuan Guan, MS[a,b], Matthew M. Ruppert, MS[a,b], Parisa Rashidi, PhD[a,c], Azra Bihorac, MD, MS[a,b,*], Tezcan Ozrazgat-Baslanti, PhD[a,b,*]

\* These authors have contributed equally.

[a] Intelligent Critical Care Center, University of Florida, Gainesville, FL

[b] Department of Medicine, Division of Nephrology, Hypertension, and Renal Transplantation, University of Florida, Gainesville, FL.

[c] Department of Biomedical Engineering, University of Florida, Gainesville, FL.

**Corresponding author:** Azra Bihorac MD MS, Department of Medicine, Division of Nephrology, Hypertension, and Renal Transplantation, PO Box 100224, Gainesville, FL 32610-0224.

Telephone: (352) 294-8580; Fax: (352) 392-5465; Email: abihorac@ufl.edu

Reprints will not be available from the author(s).

**ABSTRACT**

**Objectives:** We hypothesize that multistate models are beneficial in analyzing transitions through kidney states and understanding the underlying processes influencing the course of kidney health. Specifically, we aim to quantify longitudinal acute kidney injury (AKI) trajectories and to describe transitions through progressing and recovery states and outcomes among hospitalized patients.

**Methods:** In this large, longitudinal cohort study, 138,449 adult patients admitted to a quaternary care hospital between January 2012 and August 2019 were staged based on Kidney Disease: Improving Global Outcomes (KDIGO) serum creatinine criteria as No AKI, Stage 1, Stage 2, Stage 3, and Stage 3 with renal replacement therapy (RRT) AKI for the first 14 days of their hospital stay. We fit and examined multistate models to estimate probability of being in a certain clinical state at a given time after entering each one of the AKI stages. We investigated the effects of age, sex, race, admission comorbidities, and prolonged intensive care unit (ICU) stay on transition rates via Cox proportional hazards regression models.

**Results:** Twenty percent of hospitalized encounters (49,325/246,964) had AKI; among patients with AKI, 66% (n = 32,739) had Stage 1 AKI, 18% (n = 8,670) had Stage 2 AKI, and 17% (n = 7,916) had AKI Stage 3 with or without RRT. At seven days following Stage 1 AKI, 69% (95% confidence interval [CI]: 68.8%-70.5%) were either resolved to No AKI or discharged, while smaller proportions of recovery (26.8%, 95% CI: 26.1%-27.5%) and discharge (17.4%, 95% CI: 16.8%-18.0%) were observed following AKI Stage 2. At fourteen days following Stage 1 AKI, patients with more frail conditions (Charlson comorbidity index ≥ 3 and had prolonged ICU stay) had lower proportion of transitioning to No AKI or discharge states, while No AKI patients with these two conditions had higher probability of either developing AKI or death.

**Discussion**: Development of AKI with various severities occurs during early periods of the patients' hospitalization. Multistate analyses showed that the majority of Stage 2 and higher severity AKI patients could not resolve within seven days; therefore, strategies preventing the persistence or progression of AKI would contribute to the patients' life quality.

**Conclusions:** It is essential to identify patients at risk of developing AKI at any stage to guide treatment-related decisions. We demonstrate multistate modeling framework's utility as a mechanism for a better understanding of the clinical course of AKI with the potential to facilitate treatment and resource planning.

**Funding Sources:** This research was supported by K01 DK120784 from the National Institute of Diabetes and Digestive and Kidney Diseases (NIH/NIDDK)).

**Keywords:** Multistate, acute kidney injury, transition, longitudinal, electronic health records, hazard

**INTRODUCTION**

Acute kidney injury (AKI) occurs in almost 25% of hospitalized patients and up to 60% of patients in the intensive care unit (ICU).[1-3] Persistence of AKI or insufficient recovery of renal function exacerbates risk for chronic critical illness with reduced long-term survival and life quality.[4, 5] To optimize and tailor  clinical actions and their timely delivery, it is imperative to understand clinical course of AKI in terms of severity and recovery during hospitalization.

Conventional survival analysis methods have been utilized to describe AKI trajectories and associated outcomes.[4, 6, 7] These models have the capacity to deal with time-to-event type data and censored subjects, where a subject being censored refers to failing to experience the study's event of interest or being dropped out of the study by the end of the observation period or follow-up time.[8, 9]  Despite its significant merits, traditional survival analysis methods, such as Kaplan-Meier methods, have certain limitations. For instance, censoring action used in these models could be considered as uninformative since in real-world scenarios patients are subject to several competing risks. Competing risks models could deal with aforementioned structures; however, both approaches treat all states as *absorbing* and therefore lack inclusion of patients' history. For analyses that involve patient histories consisting of experiencing several events of interest, multistate models could be used to characterize the competing risks. Applications of multistate models could be found for various care levels[10, 11] and different patient groups such as kidney disease[12], diabetic[13], surgical[14], cancer[15], COVID-19[16], and geriatric[17] cohorts .

Multistate models are specifically beneficial in analyzing temporal changes and present an alternative approach with considerable potential in research studies with longitudinal nature; however, multistate models require precise and detailed records of transitions between the identified states. In that regard, we performed a retrospective, longitudinal study on a large cohort consisting of 242,643 adult hospitalizations, 48,915 of which were diagnosed with AKI based on KDIGO criteria.[18] Our objectives in this study could be summarized as: 1) to understand the clinical course of AKI among hospitalized adult patients by estimating probability of being in a certain clinical state at a given time after entering each one of the AKI stages, and 2) to investigate the effects of age, sex, race, comorbidities, and prolonged ICU stay on transition rates via Cox proportional hazards regression models.

## METHODS

### Study design

The study was designed and approved by the Institutional Review Board of the University of Florida and the University of Florida Privacy Office (IRB 201901123). The University of Florida Health (UFH) Integrated Data Repository acted as Honest Broker, a single-center, longitudinal dataset was curated from the electronic health records of 156,699 adult patients admitted to UFH between January 1, 2012, and August 22, 2019. We excluded patients with end stage renal disease (ESRD), encounters with no serum creatinine measurement to determine AKI status during hospitalization and within 48 hours of hospital admission, and encounters discharged within 24 hours of admission. Our final cohort included 245,663 hospital encounters from 128,271 patients (Supplemental Figure 1, Supplemental Methods).

### Assessment of kidney function and study outcomes

In identifying and staging the AKI, we used a validated computable phenotyping algorithm[19] that relies on Kidney Disease: Improving Global Outcomes (KDIGO) serum creatinine criteria.[18, 20, 21]  Reference creatinine was determined using preadmission records[22] or estimated using the Chronic Kidney Disease Epidemiology Collaboration (CKD-EPI) Study equation refit without race multiplier, as per recommendations, with a baseline estimated glomerular filtration rate (eGFR) assumption of 75 ml/min/per 1.73 m$^2$ (Supplemental Methods).[18, 23-25] We identified primary clinical outcomes as No AKI, AKI Stage 1, AKI Stage 2, AKI Stage 3, AKI Stage 3 with RRT, hospital death, and discharge. Details regarding

assumptions and the phenotyping algorithm pipeline can be found in Ozrazgat-Baslanti et al. (2019).[19]

**Multistate analyses**

Multistate models allow intermediate events to change the risk of reaching a terminal state simultaneously.[13] In defining and fitting a multistate model, two main component sets are required: state set and transition set. In this context, state set represents temporal status of a patient whereas transition set defines possible movements between states. States could be categorized as absorbing (or terminal) if leaving that particular state is impossible. In other words, if a subject moves into an absorbing state, the subject stays in that state forever. On the contrary, a state is considered transient if moving from that state or transitioning to another one is possible. Basically, state set of a multistate model is a collection of initial state(s), transient state(s), and terminal state(s). Initial states could be considered as the time point of the subject's entry into the model and returning to an initial state is not possible after the state was left. Apart from absorbing and initial states, remaining transient states could be visited several times. The collection of states and transitions presents the framework for designating a statistical model for hazard function for each of the transition identified.

Multistate models assist in quantifying separate transition intensities for switching from one particular state to another state and in quantifying the present proportion of the patients occupying a specified state at a given time point. Therefore, these models allow estimating the probability of a clinical event occurring after an entrance to a particular state over an extended time. We refer the reader for relevant background of the non-parametric or semi-parametric models to Andersen et al.[26], Thernau et al.[27], and Geskus et al. [28].

We developed two separate multistate models by using a large dataset that assembles both time-varying and static information of the patients. We identified eight mutually exclusive states based on patients' clinical condition at each time point. These states are enumerated and listed as follows: 0) Admission, 1) No AKI, 2) AKI Stage 1, 3) AKI Stage 2, 4) AKI Stage 3, 5) AKI Stage 3 with RRT, 6) Death, and 7) Discharge (Figure 1A). States were discretely determined by considering the worst AKI condition a subject experienced within 24-hour time periods.

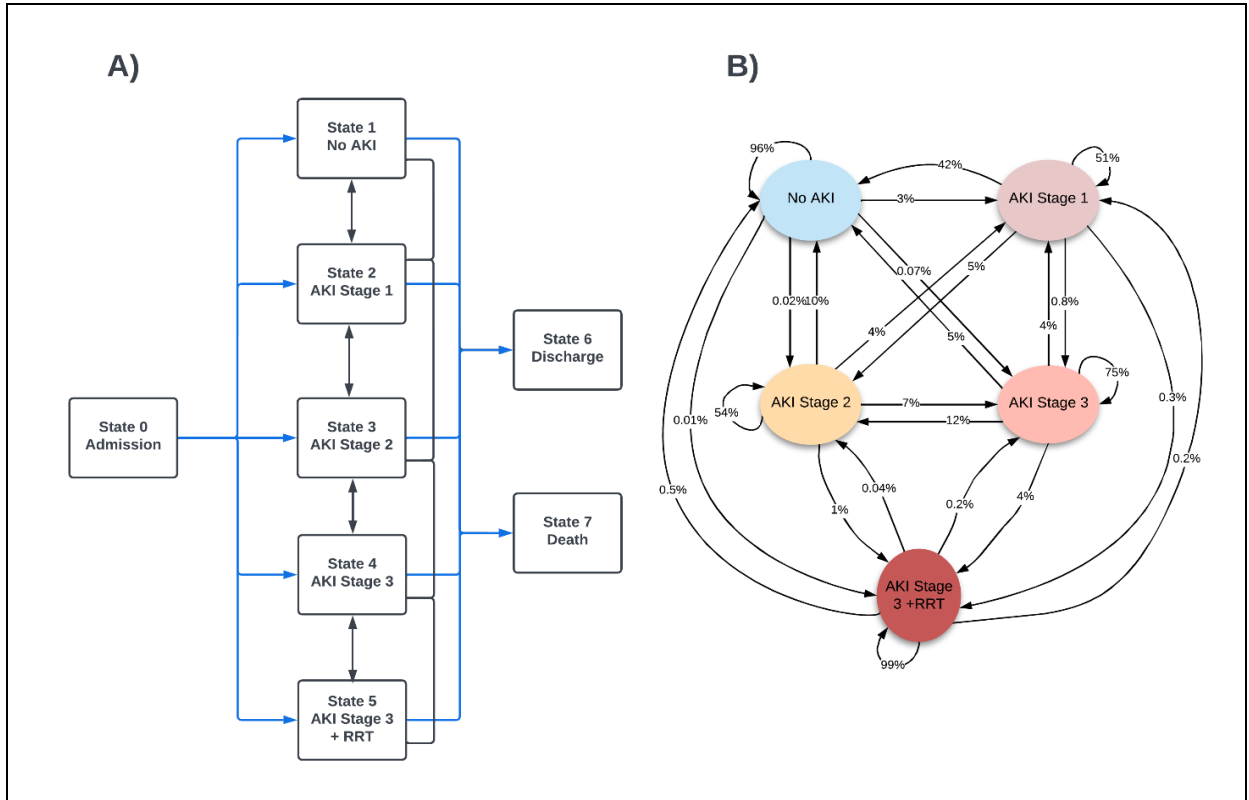

**Figure 1:** State space diagram for clinical states used in multistate analysis (A) and daily transitions in the cohort (B) (Black and blue connections in left panel indicate unidirectional and bidirectional relations, respectively.)

We first quantified the transition probabilities using an Aalen-Johansen estimation-based non-parametric multistate model where the covariate effects were ignored in estimating transition probabilities from one state to another. Following that estimate, in describing the covariate effects on the hazards, we fit a Cox model in the multistate semi-parametric framework. This approach aids in specifying distinct covariate effects for distinct transitions in the terminal states either with or without inclusion of intermediate events. We included age, sex, race, Charlson comorbidity index (CCI), and prolonged ICU stay as the covariates.[29] We presented the clinical course of the AKI patients via alluvial plots where we stratified the patients by considering their movements between the specified states throughout their hospitalization (Figure 1B).[30]

We calculated instantaneous hazard rates for particular events from predetermined states without considering intermediate events. Specifically, the instantaneous hazard rates from No AKI, AKI Stage 1, AKI Stage 2, and AKI Stage 3 states to these predetermined states including death state were calculated. For all analyses performed, we considered time 0 as the entrance moment into a certain state. Patients were censored at hospital discharge, death, or at the end of 14-day of observation period, whichever came first. Patient characteristics were presented in terms of their means and standard deviations (SD), medians with interquartile ranges (IQR), or frequencies with percentages, as appropriate. Kruskall-Wallis test or chi-square tests were performed to compare the data between groups, where appropriate. Threshold was set to 0.05 for a p-value to indicate statistical significance. All data processing and analyses were performed using Python 3.8 and R 4.1.2. We conducted modeling and statistical analyses using mstate and survival packages.[31-34]

**RESULTS**

**Clinical characteristics of patients**

Subjects were categorized with respect to the worst AKI severity outcome during their hospitalization. Cohort characteristics and their statistical comparisons were reported in Table 1. Average age for No AKI patients was 55 and was significantly lower than AKI groups (Table 1). Female and male sex were approximately equally distributed in patient groups, except AKI Stage 3 with RRT group for which average CCI was the highest. Hospital length of stay was significantly lower for No AKI cohort. Similarly, time period with mechanical ventilation and ICU length of stay was lower for No AKI group.

**Table 1:** Cohort characteristics and outcomes

| Features | No AKI | Stage 1 | Stage 2 | Stage 3 | Stage 3 + RRT |
|---|---|---|---|---|---|
| Number of patients, n | 111,860 | 24,358 | 7,579 | 5,458 | 1,664 |
| Number of encounters, n | 197,639 | 32,739 | 8,670 | 6,232 | 1,684 |
| Age in years, mean (SD) | 55 (18) | 60 (17)* | 60 (16)* | 59 (16)* | 59 (15)* |
| Female, n (%) | 103,864 (53) | 16,117 (49)* | 4,476 (52) | 2,919 (47)* | 657 (39)* |
| Ethnicity, n (%) | | | | | |
| Non-Hispanic | 18,298 (95) | 31,219 (95)* | 8,299 (96)* | 5,946 (95) | 1,572 (93) |
| Hispanic | 8,278 (4) | 1,190 (4)* | 277 (3)* | 207 (3)* | 61 (4) |
| Missing | 2,064 (1) | 331 (1) | 94 (1) | 80 (1) | 51 (3)* |
| Race, n (%) | | | | | |
| White | 144,722 (73) | 23,386 (71)* | 6,265 (72) | 4,258 (68)* | 1,195 (71) |
| African American | 40,028 (20) | 7,432 (23)* | 1,904 (22)* | 1,619 (26)* | 348 (21) |
| Other | 10,857 (5) | 1,580 (5)* | 406 (5)* | 286 (5)* | 86 (5)* |
| Missing | 2,033 (1) | 342 (1) | 95 (1) | 70 (1) | 55 (3)* |
| Smoking Status, n (%) | | | | | |
| Never | 79,680 (40) | 12,053 (37)* | 3,152 (36)* | 2,289 (37)* | 547 (32)* |
| Former | 58,823 (30) | 11,530 (35)* | 2,916 (34)* | 2,017 (32)* | 581 (35)* |
| Current | 42,289 (21) | 5,967 (18)* | 1,634 (19)* | 1,194 (19)* | 260 (15)* |
| Missing | 16,848 (9)* | 3,190 (10)* | 968 (11)* | 733 (12)* | 296 (18)* |
| CKD, n (%) | | | | | |
| CKD | 33,666 (17) | 9,798 (30)* | 2,350 (27)* | 2,187 (35)* | 449 (27)* |
| No CKD | 90,093 (46) | 12,635 (39)* | 3,732 (43)* | 2,268 (36)* | 418 (25)* |
| Missing | 73,881 (37) | 10,307 (31)* | 2,588 (30)* | 1,778 (29)* | 817 (49)* |
| Comorbid conditions on admission date, n (%) | | | | | |
| Charlson comorbidity index, median (IQR) | 2 (0, 5) | 3 (1, 7)* | 3 (1, 6)* | 4 (1, 7)* | 2 (0, 6) |
| Charlson comorbidity index ≥ 3, n (%) | 51,680 (26) | 14,938 (46)* | 4,065 (47)* | 3,296 (53)* | 1,052 (62)* |
| Myocardial Infarction | 23,048 (12) | 5,770 (18)* | 1,337 (15)* | 1,038 (17)* | 226 (13) |
| Congestive Heart Failure | 39,098 (20) | 10,687 (33)* | 2,749 (32)* | 2,091 (34)* | 484 (29)* |
| Peripheral Vascular Disease | 38,326 (19) | 9,734 (30)* | 2,440 (28)* | 1,704 (27)* | 445 (26)* |
| Diabetes | 49,025 (25) | 11,848 (36)* | 30,44 (35)* | 2,269 (36)* | 477 (28)* |
| Hypertension | 30,587 (15) | 9,099 (28)* | 2,000 (23)* | 1,957 (31)* | 403 (24)* |
| Cancer | 43,129 (22) | 7,825 (24)* | 2,153 (25)* | 1,612 (26)* | 291 (17)* |
| Metastatic Carcinoma | 14,753 (7) | 2,567 (8) | 799 (9)* | 660 (11)* | 67 (4)* |
| Moderate-Severe Liver Disease | 37,569 (19) | 7,441 (23)* | 2,275 (26)* | 1,684 (27)* | 425 (25)* |
| Reference creatinine, mean (SD) | 0.84 (0.33) | 0.95 (0.46)* | 0.83 (0.32) | 1.30 (1.29)* | 1.22 (0.98)* |

| Features | No AKI | Stage 1 | Stage 2 | Stage 3 | Stage 3 + RRT |
|---|---|---|---|---|---|
| Mechanical ventilation days, median (IQR) | 0.00 (0.00, 0.00) | 0.00 (0.00, 0.00)* | 0.00 (0.00, 0.46)* | 0.00 (0.00, 0.79)* | 5.26 (1.33, 13.57) * |
| Length of stay (days), median (IQR) | 3.29 (2.01, 5.71) | 6.62 (3.67, 12.00)* | 8.09 (4.18, 15.33)* | 8.60 (4.53, 16.78)* | 21.60 (11.13, 36.52)* |
| ICU length of stay (days), median (IQR) | 0.00 (0.00, 0.30) | 0.00 (0.00, 4.25)* | 1.88 (0.00, 6.90)* | 2.26 (0.00, 7.53)* | 13.91 (5.19, 27.44)* |
| ICU length of stay ≥ 48 hours, n (%) | 29,609 (15) | 12,308 (38)* | 4,250 (49)* | 3,256 (52)* | 1,489 (88)* |
| Hospital mortality, n (%) | 2,357 (1) | 1,560 (5)* | 949 (11)* | 1,118 (18)* | 831 (49)* |

Star sign (*) in Table 1 indicates a Bonferroni corrected p-value ≤ 0.05 compared to no AKI group.

**AKI trajectory analyses**

Figure 2 demonstrates daily transitions in the cohorts for each AKI stage. Patients admitted with AKI Stage 1 had the highest proportion for early resolution or discharge with the lowest percentage for AKI progression (Figure 2 A). Patients admitted with worse AKI severity had almost consistently lower proportions for resolution and discharge compared to AKI Stage 1 group. Similarly, patients with more severe AKI groups on their early days of hospitalization had higher percentages for maintaining their initial AKI stage (Figure 2 B-D).

Within 24 hours following the admission, 8.1% of the patients had AKI, a majority of which had Stage 1 AKI (Figure 3, Table 2). From seven days after admission, 3.95% (95% CI, 3.87%-4.02%) of the cohort had Stage 1 AKI whereas 2.12% (95% CI, 2.05%-2.20%) experienced Stage 2 or more severe AKI. At that point, 35.07% (95% CI, 34.90%-35.25%) of the cohort were discharged and the probability for terminal state of death was 0.39% (95% CI, 0.37%-0.42%). On day 7 following AKI Stage 2, proportion of progression to higher stages of AKI (3.84% (95% CI, 3.30%-5.85%)) was higher than the proportion of progression from AKI Stage 1 (3.45% (95% CI, 3.20%-3.72%)).

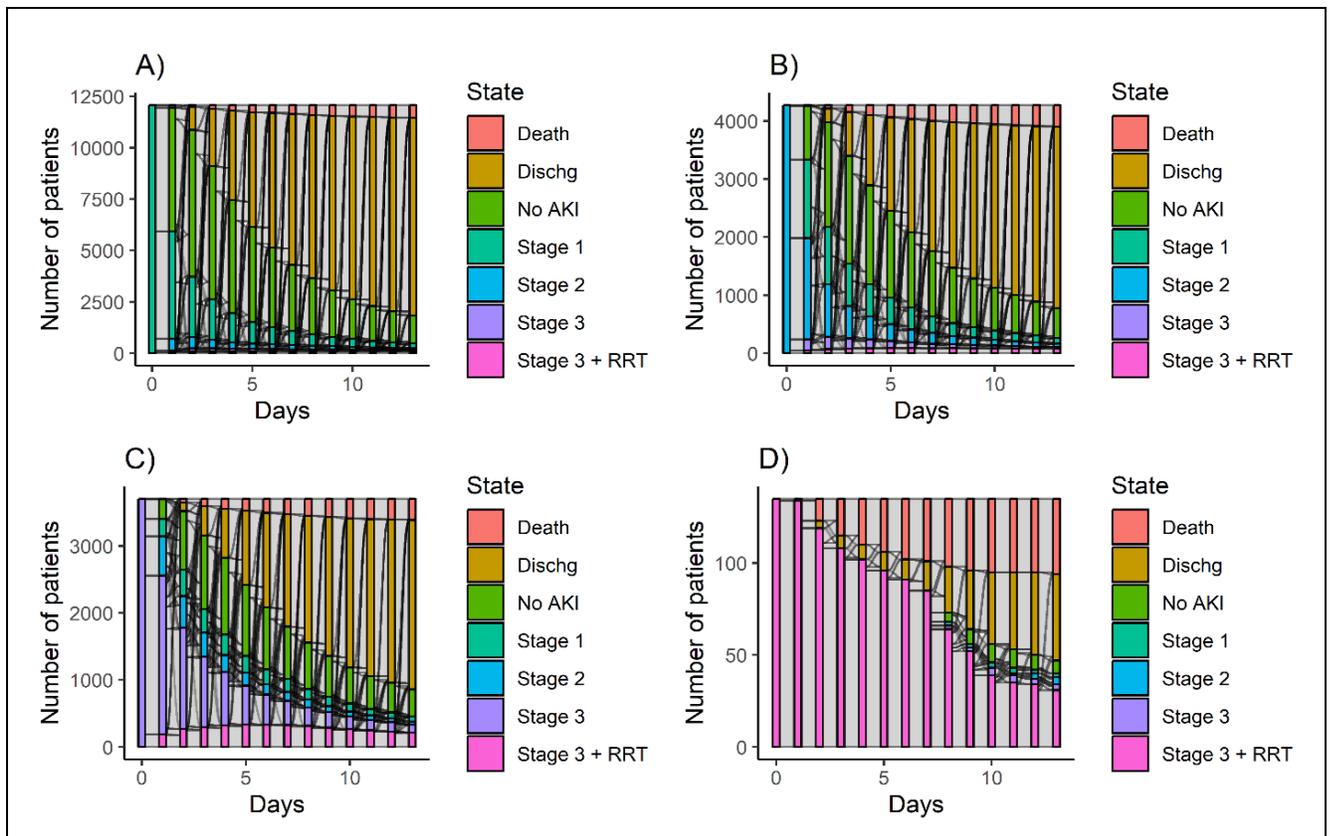

**Figure 2:** Number of patients transitioning on consecutive days shown for 14 days of hospitalization for AKI Stage 1 (A), AKI Stage 2 (B), AKI Stage 3 (C), and AKI Stage 3 with RRT (D) patients. The horizontal axis of the figures in the panels represent the days in hospital and the vertical axis displays the number of patients.

At that time point, resolved or discharged percentage of the patients with an initial status of AKI Stage 1 was the highest (69.68% (95% CI, 68.86%-70.51%)) than cohorts with AKI Stage 2, 3 and 3 with RRT. Among AKI patients without RRT requirement, Stage 3 patients has the highest percentage for persisting condition (53.90% [95% CI, 52.46%-55.37%]).

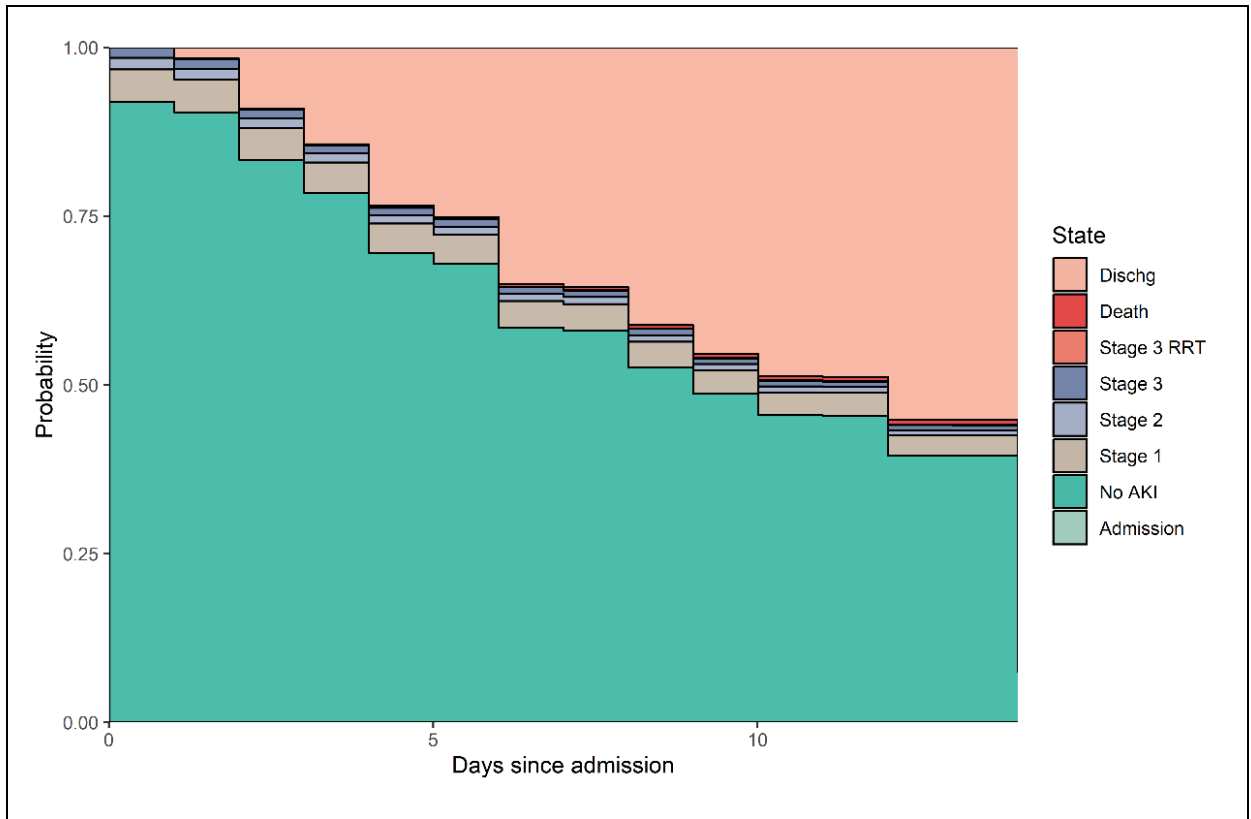

**Figure 3:** Estimated proportions of the patients in each state on any day following admission.

Patients with No AKI diagnosis had the highest transition rates to AKI Stage 1 state throughout the hospitalization (Figure 4 A). For those patients, first peak was observed between day six and day seven, whereas the second peak transfer rate to AKI Stage 1 was observed near end of the 14-day hospital stay. Transfer rates to death repeated a similar pattern of peaks and lows with transfer hazards to AKI Stage 1 with lower rates. AKI Stage 1 patients were more likely to start resolving between day two and three (Figure 4 B). Among AKI Stage 1 patients, top transfer rates occurred for resolving state and the second top transfer rates indicated to next advanced AKI stage which is Stage 2. These two top processes had slightly different timings for jumps occurred in a 14-day time period.

**Table 2:** Multistate models based estimated proportions of patients in each clinical state over time

| Patient Characteristics and Outcomes | No AKI, % (95% CI) | AKI Stage 1, % (95% CI) | AKI Stage 2, % (95% CI) | AKI Stage 3, % (95% CI) | AKI Stage 3 + RRT, % (95% CI) | Death, % (95% CI) | Discharge, % (95% CI) |
|---|---|---|---|---|---|---|---|
| **Days since admission** | | | | | | | |
| Day 1 | 91.91 (91.91, 91.91) | 4.84 (4.84, 4.84) | 1.71 (1.71, 1.71) | 1.51 (1.51, 1.51) | 0.03 (0.03, 0.03) | 0 (0, 0) | 0 (0, 0) |
| Day 3 | 83.36 (83.23, 83.48) | 4.70 (4.63, 4.76) | 1.46 (1.43, 1.49) | 1.32 (1.30, 1.34) | 0.04 (0.04, 0.05) | 0.06 (0.05, 0.07) | 9.06 (8.95, 9.17) |
| Day 7 | 58.46 (58.28, 58.65) | 3.95 (3.87, 4.02) | 1.06 (1.03, 1.10) | 0.98 (0.95, 1.01) | 0.08 (0.07, 0.09) | 0.39 (0.37, 0.42) | 35.07 (34.90, 35.25) |
| **Days since AKI Stage 1** | | | | | | | |
| Day 1 | 0 (0, 0) | 100 (100, 100) | 0 (0, 0) | 0 (0, 0) | 0 (0, 0) | 0 (0, 0) | 0 (0, 0) |
| Day 3 | 24.9 (24.27, 25.55) | 68.01 (67.29, 68.73) | 1.80 (1.63, 1.99) | 0.18 (0.13, 0.25) | 0.07 (0.04, 0.12) | 0.18 (0.14, 0.25) | 4.86 (4.59,5.14) |
| Day 7 | 42.52 (42.09, 42.96) | 25.94 (25.44, 26.45) | 2.69 (2.54, 2.84) | 0.60 (0.53, 0.67) | 0.16 (0.13, 0.21) | 0.93 (0.84, 1.03) | 27.16 (26.77,27.55) |
| **Days since AKI Stage 2** | | | | | | | |
| Day 1 | 0 (0,0) | 0 (0,0) | 100 (100, 100) | 0 (0,0) | 0 (0,0) | 0 (0,0) | 0 (0,0) |
| Day 3 | 12.03 (11.26, 12.85) | 13.04 (12.25,13.89) | 70.65 (69.45, 71.88) | 1.36 (1.11, 1.66) | 0.12 (0.06, 0.24) | 0.38 (0.26, 0.55) | 2.42 (2.11, 2.77) |
| Day 7 | 26.81 (26.11, 27.53) | 18.62 (17.99,19.26) | 31.15 (30.11, 32.22) | 3.33 (3.01, 3.68) | 0.51 (0.39, 0.67) | 2.17 (1.90, 2.46) | 17.42 (16.85, 18.01) |
| **Days Since AKI Stage 3** | | | | | | | |
| Day 1 | 0 (0,0) | 0 (0,0) | 0 (0,0) | 100 (100, 100) | 0 (0,0) | 0 (0,0) | 0 (0,0) |
| Day 3 | 4.78 (4.21, 5.42) | 3.09 (2.65,3.60) | 5.86 (5.22,6.58) | 83.97 (82.85, 85.10) | 0.43 (0.28, 0.67) | 0.37 (0.23, 0.58) | 1.51 (1.21, 1.87) |
| Day 7 | 12.99 (12.28, 13.74) | 7.27 (6.78,7.80) | 11.11 (10.39,11.88) | 53.90 (52.46, 55.37) | 1.49 (1.20, 1.85) | 2.89 (2.48, 3.35) | 10.36 (9.71, 11.04) |
| **Days since AKI Stage 3 with RRT** | | | | | | | |
| Day 1 | 0 (0,0) | 0 (0,0) | 0 (0,0) | 0 (0,0) | 100 (100,100) | 0 (0,0) | 0 (0,0) |
| Day 3 | 0 (0,0) | 0 (0,0) | 0 (0,0) | 0 (0,0) | 90.36 (84.52, 96.60) | 6.02 (2.73, 13.30) | 3.61 (1.30,13.30) |
| Day 7 | 0 (0,0) | 0 (0,0) | 0 (0,0) | 0 (0,0) | 69.81 (62.86, 77.52) | 19.90 (14.56, 27.18) | 10.30 (6.32, 16.78) |

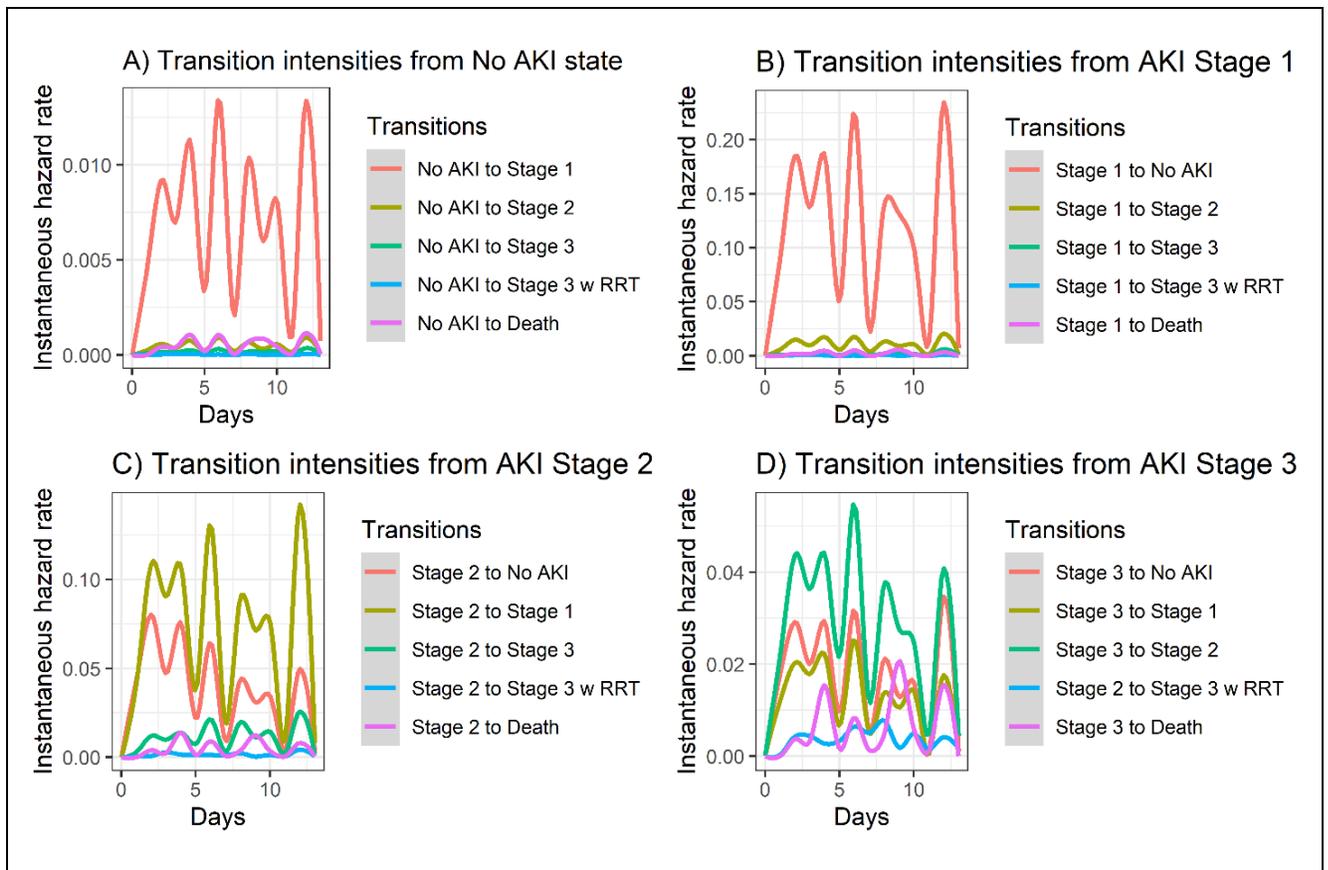

**Figure 4:** Instantaneous hazard rates of outcomes from clinical states in hospitalized patients based on multistate analysis.

AKI Stage 2 cohort had the highest transition rates for regression to AKI Sage 1 and resolution (Figure 4 C). Among patients with AKI Stage 2, hazard rates for transitioning to AKI Stage 1 were higher for the first seven days, and followed an irregular path thereafter with several declines and a peak near completion of two-week length of stay. Apart from that irregularity, risk for progressing to AKI Stage 3 was elevated in the period between day five and day ten. AKI Stage 3 patients were more likely to transfer to AKI Stage 2 compared to transitions to AKI Stage 1 and No AKI states (Figure 4 D). Transition rates from AKI Stage 3 to AKI Stage 2, AKI Stage 1, and No AKI were at their highest around day six. Competing risk for transfer to death reached top between day eight and nine.

We fit a Cox model that involves transition-specific covariates for each specified transfer form a given to another. In that model, the patient's age, sex (female vs male), race (African-American vs non–African American), CCI, and prolonged ICU stay were included. Age was dichotomized as age < 65 and age ≥ 65 years old and CCI was dichotomized as CCI < 3 and CCI ≥ 3. Prolonged ICU was indicated with a length of stay longer than 48 hours in ICU. We reported the Cox regression model coefficients and p-values for these covariates in Supplementary Table 1. According to the results, these covariates were statistically significant in a majority of the identified transitions.

Estimated percentages of patients for each state at any given day during hospital stay were mainly different with respect to their accompanied admission comorbidities and need of prolonged ICU stay (Figure 5, 6, Supplemental Figures 2, 3, 4). Among AKI Stage 1 cohort, patients accompanied with the most severe conditions (i.e., CCI ≥ 3 and ICU length of stay ≥ 48 hours) had greater proportion of patients for sustained AKI Stage 1 severity and progression to AKI Stage 2 and AKI Stage 3 and lower discharge percentage compared to patients with milder conditions (i.e., CCI < 3 and ICU length of stay < 48 hours) (Figure 5 and 6). Proportion of transfer to death condition from AKI Stage 1 were more pronounced among male patients with age ≥ 65 near the end of the 14-day period of hospital stay. Considering the cohorts with more frail conditions, percentage of AKI Stage 3 progression was slightly higher for African-American male patients compared to non–African American subjects. In addition, similar to non-parametric analyses presented in this study, more advanced AKI stages were observed with higher tendency towards either maintaining current AKI condition or regressing to its neighbor AKI Stage compared to AKI Stage 1 group (Supplementary Figures 2-4).

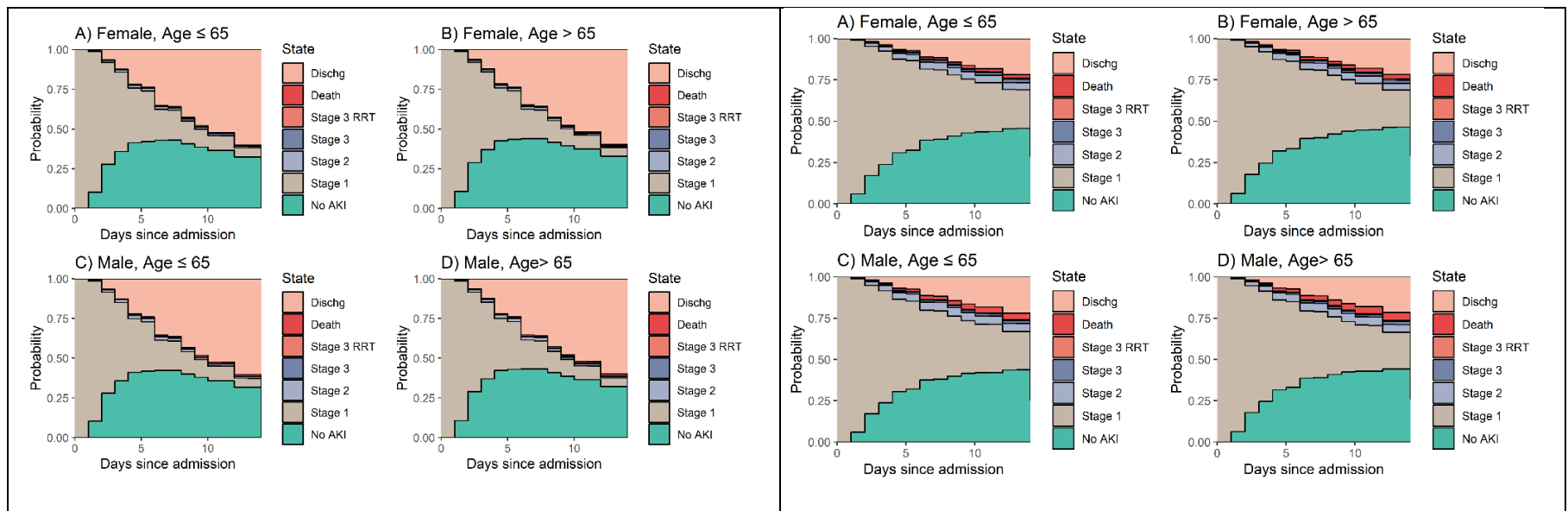

**Figure 5:** Proportion of non–African American patients estimated to be in each clinical state for AKI Stage 1 patients with CCI < 3 and ICU < 48 hours (left panel) and AKI Stage 1 patients with CCI ≥ 3 and ICU stay ≥ 48 hours (right panel) for 14 days.

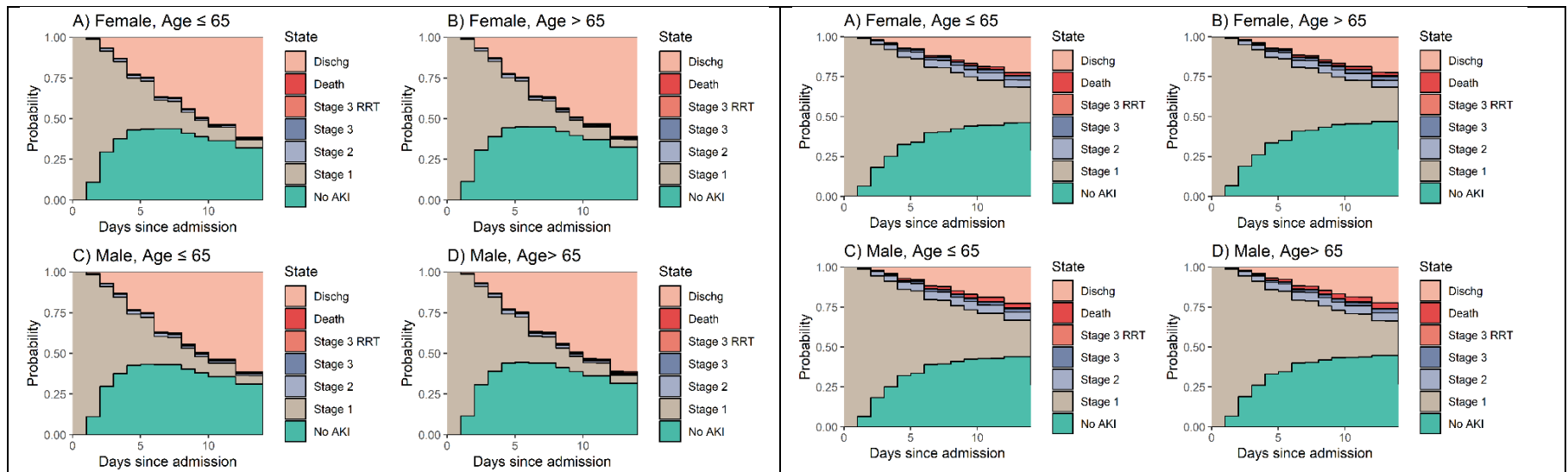

**Figure 6:** Proportion of African American patients estimated to be in each clinical state for AKI Stage 1 patients with CCI < 3 and ICU < 48 hours (left panel) and AKI Stage 2 patients with CCI ≥ 3 and ICU stay ≥ 48 hours (right panel) for 14 days.

**DISCUSSION**

We considered a large cohort of hospitalized patients and retrospectively characterized subjects' AKI trajectory by using multistate models that consider the cohort's longitudinal outcomes. Specifically, we fit multistate models for a state-space framework that indicates the clinical status in addition to feasible transitions between them, and we described the transition dynamics in terms of hazard rates and transition probabilities.[35] In this cohort, any stage AKI was developed among 20% of the patients, where the majority of AKI outcome was labeled as Stage 1. Tabulated admission comorbidity frequencies were higher among patients with AKI compared to No AKI stratum. Specifically, CCI value greater than three was associated with more advanced AKI stage, where there were significant and gradual increases in its occurrence for patients with No AKI (26%), AKI Stage 1 (46%), AKI Stage 2 (47%), AKI Stage 3 (53%), and AKI Stage 3 with RRT (62%). Additionally, more advanced AKI trajectories were also associated with longer mechanical ventilation period, ICU stay, and hospitalization.

In order to understand the transition processes between clinical states regarding the AKI status, death, and discharge, we modeled the patients' experiences over the course of the first 14 days of hospitalization. Towards that aim, we first calculated Aalen-Johansen estimators via non-parametric multistate models and estimated the instantaneous hazard rates of each transition occurrence in addition to the probability of being in certain states at a given time point. This granular and time-dependent analyses demonstrated AKI Stage 1 as the AKI condition with the lowest tendency towards developing more advanced stages. To clarify, after seven days following initiation of AKI Stage 1, 69% of this patient group had either resolved AKI or were discharged. In contrast, seven days following AKI Stage 2, estimated proportion of patients either being discharged or with no AKI was 33.23%, and if regression to AKI Stage 1 was also included, the outcome with more desirable conditions increased to 64.38%.

We expanded the initial multistate analysis by modeling transitions' hazard rates with Cox-type regression models to investigate the effect of a set of covariates that are potentially associated with events of interest. According to Cox-type models, the covariates indicate the severity of the overall condition of the subjects (i.e., CCI and prolonged ICU stay) heavily influenced the probability of being at a resolving or discharge. To clarify, near the end of a 14-day period of hospitalization with a Stage 1 AKI initiated at the beginning, patients admitted with a higher number of comorbidities with prolonged ICU stay had higher percentage for being at AKI Stage 1 state and transitioning to either progressed AKI stages or death states. In line with that, No AKI patients admitted with a higher number of comorbidities and who stayed ICU longer than 48 hours had higher probability of being in either AKI stage or death.

The motivation for performing this study was that the clinical course of AKI for hospitalized patients had not been sufficiently described with multistate models that exploit granular and longitudinal structures, despite similar work having been performed for smaller patient populations. In a retrospective analysis of critically ill COVID-19 patients, a cohort of 367 subjects were considered and their AKI transitions were described with multistate models.[36] In this study, Lyons et al. presented the estimated probabilities of being in a specified clinical status where the AKI related states rely on the worst AKI stage within 12-hour blocks. Another recent multistate modeling application for investigating kidney disease progression was given for a cohort of 225 patients who were prescribed colistin.[37] In addition to AKI, a retrospective study for modeling transitions between CKD stages via multistate methods was given for a cohort of 117 hospitalized and non-hospitalized patients.[38] To our knowledge, this study is the first large-scale, granular application of multistate methods for describing the characteristics influencing the transitions towards progressed or regressed AKI stages in addition to discharge and death states. With the aid of a large, diverse cohort of hospitalized subjects, we utilized multistate methods in modeling the longitudinal outcomes regarding the patients' AKI status.

Multistate models developed in this study output a probabilistic way to describe clinical course of AKI among hospitalized patients. Those estimations could be utilized in planning prevention decisions, resource usage, and timely intervention of AKI. Despite our use of a large and diverse cohort in building the multistate models, the cohort relies on a single institution. Therefore, this single-center design limits generalizability to other practice settings. The timing precision of a transition relies on the serum creatinine measure date and time, so a future work could be based on defining AKI conditions considering hourly basis urine output criteria, which also allows more precise and timely monitoring as well as transition time records.

**CONCLUSIONS**

Harnessing the granular and longitudinal information processing capability of multistate models, we estimated possible pathways in clinical trajectories of AKI among hospitalized patients, thus stressing the ability of this approach to convey insights into AKI course from a probabilistic perspective. Moreover, the large and diverse cohort was expected to assist mitigating the bias in fitting the model. Precise and timely identification of patients at elevated risk for AKI progress or other terminal states may facilitate the delivery of tailored treatments that prevent adverse outcomes or foster kidney recovery to improve life quality and optimize resource planning.

## ACKNOWLEDGMENTS

**Author Contributions:** Dr. Tezcan Ozrazgat-Baslanti and Dr. Azra Bihorac had full access to all the data in the study and take responsibility for the integrity of the data and the accuracy of the data analysis.

Concept and design: Bihorac, Ozrazgat-Baslanti, Adiyeke

Acquisition, analysis, or interpretation of data: Adıyeke, Ren, Guan, Ruppert, Ozrazgat-Baslanti, Bihorac

Drafting of the manuscript: Adiyeke, Ozrazgat-Baslanti, Bihorac

Critical revision of the manuscript for important intellectual content: Adiyeke, Ren, Guan, Ruppert, Rashidi, Ozrazgat-Baslanti, Bihorac

Obtained funding: Ozrazgat-Baslanti

Supervision: Ozrazgat-Baslanti, Bihorac

**Additional Contributions:** We would like to acknowledge Intelligent Critical Care Center research group. We acknowledge the University of Florida Integrated Data Repository (IDR) and the UF Health Office of the Chief Data Officer for providing the analytic data set for this project.

All authors approved the final version of the manuscript.

## COMPETING INTERESTS

# References

1. Darmon M, Ostermann M, Cerda J, Dimopoulos MA, Forni L, Hoste E, et al. Diagnostic work-up and specific causes of acute kidney injury. *Intensive Care Med*. 2017;43:829-840

2. James MT, Bhatt M, Pannu N, Tonelli M. Long-term outcomes of acute kidney injury and strategies for improved care. *Nat Rev Nephrol*. 2020;16:193-205

3. Sawhney S, Fraser SD. Epidemiology of aki: Utilizing large databases to determine the burden of aki. *Adv Chronic Kidney Dis*. 2017;24:194-204

4. Ozrazgat-Baslanti T, Loftus TJ, Ren Y, Adiyeke E, Miao S, Hashemighouchani H, et al. Association of persistent acute kidney injury and renal recovery with mortality in hospitalised patients. *BMJ Health Care Inform*. 2021;28

5. Gardner AK, Ghita GL, Wang Z, Ozrazgat-Baslanti T, Raymond SL, Mankowski RT, et al. The development of chronic critical illness determines physical function, quality of life, and long-term survival among early survivors of sepsis in surgical icus. *Crit Care Med*. 2019;47:566-573

6. Bhatraju PK, Zelnick LR, Chinchilli VM, Moledina DG, Coca SG, Parikh CR, et al. Association between early recovery of kidney function after acute kidney injury and long-term clinical outcomes. *JAMA Network Open*. 2020;3:e202682-e202682

7. Ozrazgat-Baslanti T, Loftus TJ, Mohandas R, Wu Q, Brakenridge S, Brumback B, et al. Clinical trajectories of acute kidney injury in surgical sepsis: A prospective observational study. *Ann Surg*. 2022;275:1184-1193

8. Ross-Driscoll K, Patzer RE. Competing risks and multistate models in clinical nephrology research. *Kidney Int Rep*. 2022;7:2325-2326

9. Le-Rademacher JG, Therneau TM, Ou F-S. The utility of multistate models: A flexible framework for time-to-event data. *Current Epidemiology Reports*. 2022;9:183-189

10. Ursino M, Dupuis C, Buetti N, de Montmollin E, Bouadma L, Golgran-Toledano D, et al. Multistate modeling of covid-19 patients using a large multicentric prospective cohort of critically ill patients. *J Clin Med*. 2021;10

11. Upshaw JN, Konstam MA, Klaveren Dv, Noubary F, Huggins GS, Kent DM. Multistate model to predict heart failure hospitalizations and all-cause mortality in outpatients with heart failure with reduced ejection fraction. *Circulation: Heart Failure*. 2016;9:e003146

12. Fathi M, Khakifirooz M. Kidney-related operations research: A review. *IISE Transactions on Healthcare Systems Engineering*. 2019;9:226-242

13. Wang WM, Ou HT, Wen MJ, Su PF, Yang CY, Kuo TH, et al. Association of retinopathy severity with cardiovascular and renal outcomes in patients with type 1 diabetes: A multi-state modeling analysis. *Sci Rep*. 2022;12:4177

14. Manzini G, Ettrich TJ, Kremer M, Kornmann M, Henne-Bruns D, Eikema DA, et al. Advantages of a multi-state approach in surgical research: How intermediate events and risk factor profile affect the prognosis of a patient with locally advanced rectal cancer. *BMC Medical Research Methodology*. 2018;18:23

15. Cheung LC, Albert PS, Das S, Cook RJ. Multistate models for the natural history of cancer progression. *British Journal of Cancer*. 2022;127:1279-1288

16. Mody A, Lyons PG, Vazquez Guillamet C, Michelson A, Yu S, Namwase AS, et al. The clinical course of coronavirus disease 2019 in a us hospital system: A multistate analysis. *American Journal of Epidemiology*. 2021;190:539-552

17. Neumann JT, Thao LT, Callander E, Carr PR, Qaderi V, Nelson MR, et al. A multistate model of health transitions in older people: A secondary analysis of aspree clinical trial data. *The Lancet Healthy Longevity*. 2022;3:e89-e97

18. Levin A, Stevens PE, Bilous RW, Coresh J, De Francisco AL, De Jong PE, et al. Kidney disease: Improving global outcomes (kdigo) ckd work group. Kdigo 2012 clinical practice guideline for the evaluation and management of chronic kidney disease. *Kidney international supplements*. 2013;3:1-150

19. Ozrazgat-Baslanti T, Hobson C, Motaei A, Islam R, Hashemighouchani H, Ruppert M, et al. Development and validation of computable phenotype to identify and characterize kidney health in adult hospitalized patients. *arXiv preprint*. 2019;2604673

20. Chawla LS, Bellomo R, Bihorac A, Goldstein SL, Siew ED, Bagshaw SM, et al. Acute kidney disease and renal recovery: Consensus report of the acute disease quality initiative (adqi) 16 workgroup. *Nat Rev Nephrol*. 2017;13:241-257

21. Khwaja A. Kdigo clinical practice guidelines for acute kidney injury. *Nephron Clin Pract*. 2012;120:c179-184

22. Selby NM, Hill R, Fluck RJ, Programme NHSETKA. Standardizing the early identification of acute kidney injury: The nhs england national patient safety alert. *Nephron*. 2015;131:113-117

23. Bellomo R, Ronco C, Kellum JA, Mehta RL, Palevsky P, Acute Dialysis Quality Initiative w. Acute renal failure - definition, outcome measures, animal models, fluid therapy and information technology needs: The second international consensus conference of the acute dialysis quality initiative (adqi) group. *Crit Care*. 2004;8:R204-212

24. Zavada J, Hoste E, Cartin-Ceba R, Calzavacca P, Gajic O, Clermont G, et al. A comparison of three methods to estimate baseline creatinine for rifle classification. *Nephrol Dial Transplant*. 2010;25:3911-3918

25. Inker LA, Eneanya ND, Coresh J, Tighiouart H, Wang D, Sang Y, et al. New creatinine- and cystatin c-based equations to estimate gfr without race. *N Engl J Med*. 2021;385:1737-1749

26. Andersen PK, Borgan O, Gill RD, Keiding N. *Statistical models based on counting processes*. Springer Science & Business Media; 2012.

27. Therneau TM, Grambsch PM, Therneau TM, Grambsch PM. *The cox model*. Springer; 2000.

28. Geskus RB. *Data analysis with competing risks and intermediate states*. CRC Press Boca Raton; 2016.

29. Deyo RA, Cherkin DC, Ciol MA. Adapting a clinical comorbidity index for use with icd-9-cm administrative databases. *J Clin Epidemiol*. 1992;45:613-619

30. Rosvall M, Bergstrom CT. Mapping change in large networks. *PLoS One*. 2010;5:e8694

31. de Wreede LC, Fiocco M, Putter H. The mstate package for estimation and prediction in non- and semi-parametric multi-state and competing risks models. *Comput Methods Programs Biomed*. 2010;99:261-274

32. de Wreede LC, Fiocco M, Putter H. Mstate: An r package for the analysis of competing risks and multi-state models. *Journal of Statistical Software*. 2011;38:1 - 30

33. Putter H, Fiocco M, Geskus RB. Tutorial in biostatistics: Competing risks and multi-state models. *Stat Med*. 2007;26:2389-2430

34. Therneau TM, Lumley T. Package 'survival'. *R Top Doc*. 2015;128:28-33

35. von Cube M, Schumacher M, Wolkewitz M. Basic parametric analysis for a multi-state model in hospital epidemiology. *BMC Med Res Methodol*. 2017;17:111

36. Lyons PG, Mody A, Bewley AF, Schoer M, Neelam Raju B, Geng E, et al. Multistate modeling of clinical trajectories and outcomes in the icu: A proof-of-concept evaluation of acute kidney injury among critically ill patients with covid-19. *Crit Care Explor*. 2022;4:e0784

37. Lintu M, Shreyas K, Kamath A. A multi-state model for kidney disease progression. *Clinical Epidemiology and Global Health*. 2022;13:100946

38. Grover G, Sabharwal A, Kumar S, Thakur AK. A multi-state markov model for the progression of chronic kidney disease. *Turkiye Klinikleri Journal of Biostatistics*. 2019;11

**Supplementary Online Content**

**Adiyeke E, Ren Y, Guan Z, Ruppert M, Rashidi P, Bihorac A, Ozrazgat-Baslanti T. Clinical Courses of Acute Kidney Injury in Hospitalized Patients: A Multistate Analysis**

This supplementary material has been provided by the authors to give readers additional information about their work.

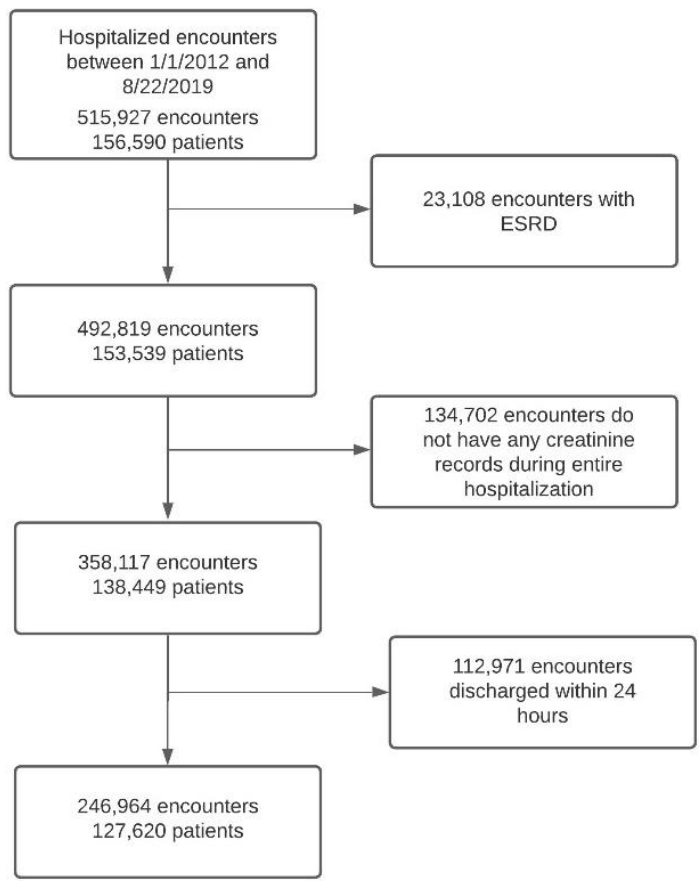

**Supplemental Figure 1.** Patient inclusion chart.

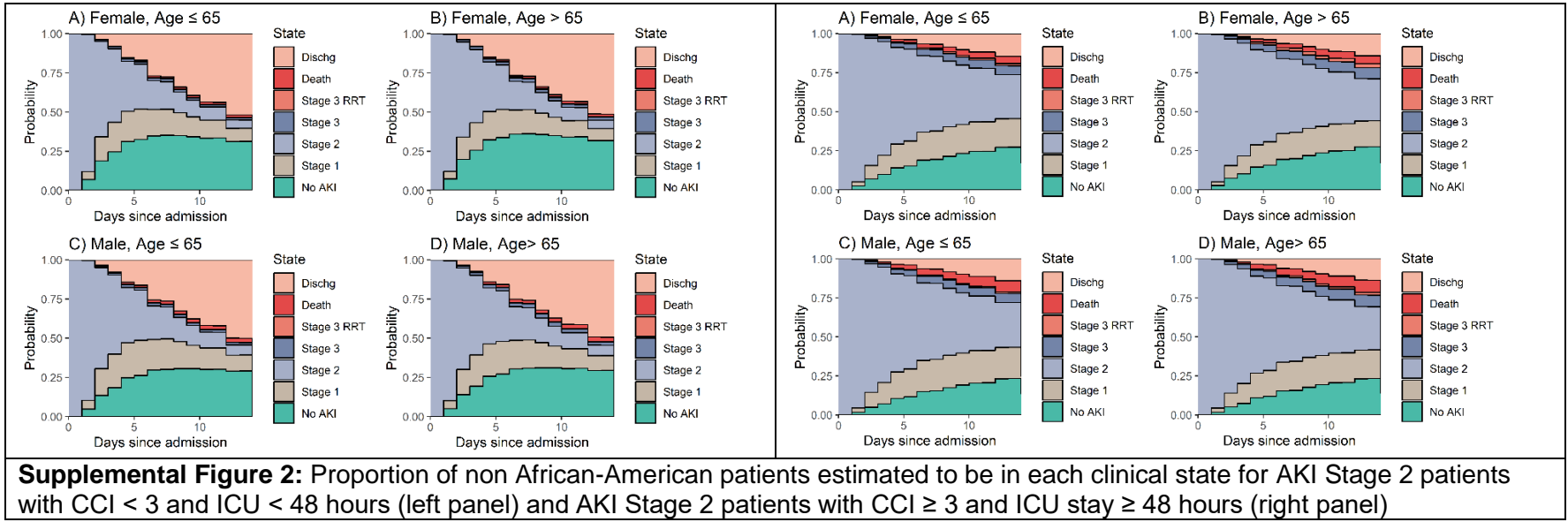

**Supplemental Figure 2:** Proportion of non African-American patients estimated to be in each clinical state for AKI Stage 2 patients with CCI < 3 and ICU < 48 hours (left panel) and AKI Stage 2 patients with CCI ≥ 3 and ICU stay ≥ 48 hours (right panel)

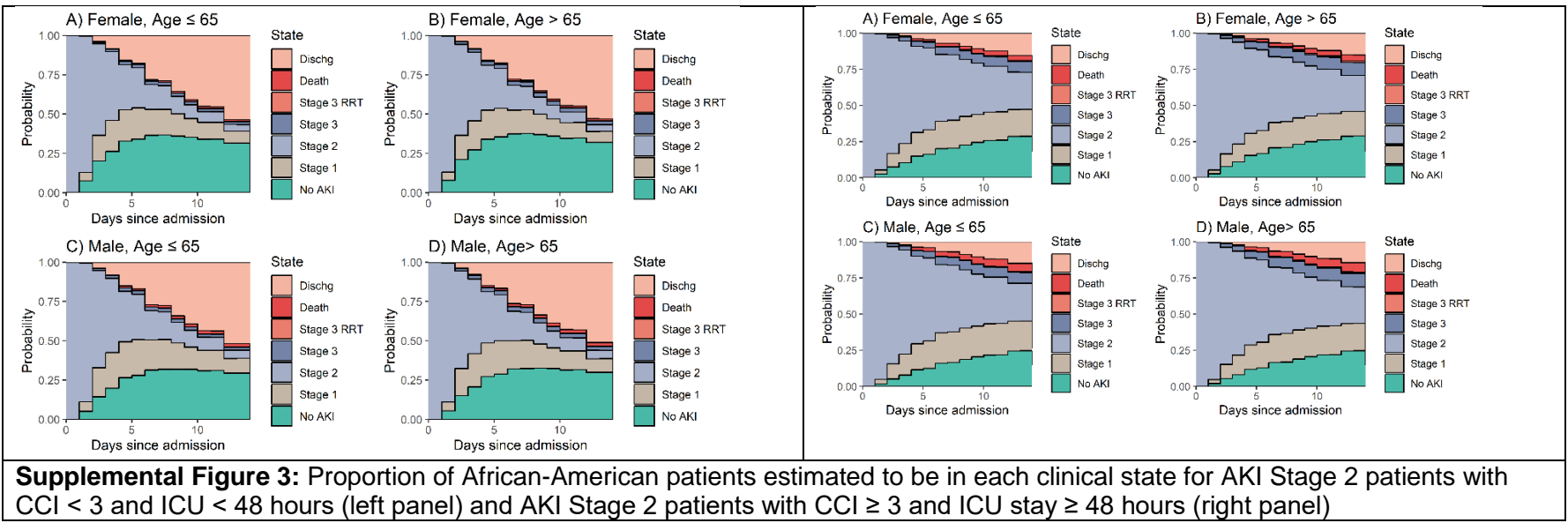

**Supplemental Figure 3:** Proportion of African-American patients estimated to be in each clinical state for AKI Stage 2 patients with CCI < 3 and ICU < 48 hours (left panel) and AKI Stage 2 patients with CCI ≥ 3 and ICU stay ≥ 48 hours (right panel)

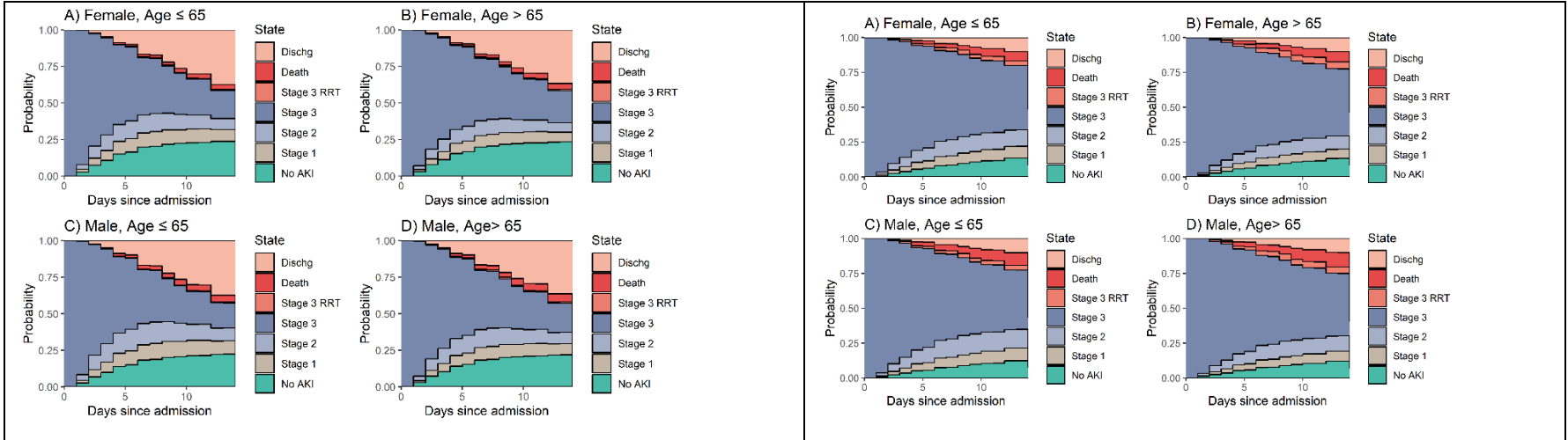

**Supplemental Figure 4:** Proportion of non African-American patients estimated to be in each clinical state for AKI Stage 3 patients with CCI < 3 and ICU < 48 hours (left panel) and AKI Stage 3 patients with CCI ≥ 3 and ICU stay ≥ 48 hours (right panel)

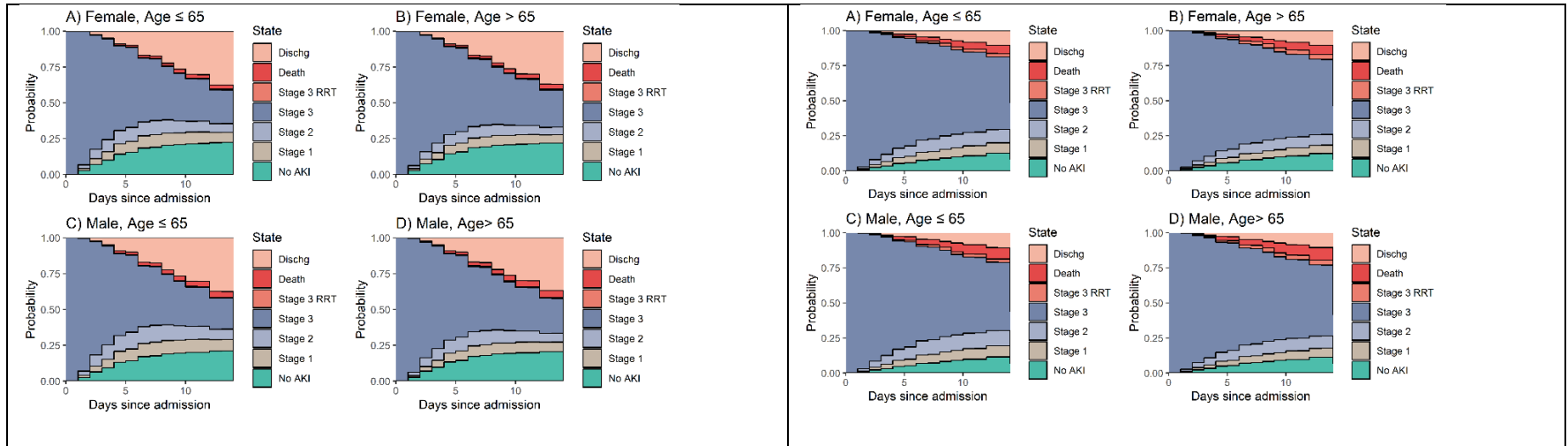

**Supplemental Figure5:** Proportion of African-American patients estimated to be in each clinical state for AKI Stage 3 patients with CCI < 3 and ICU < 48 hours (left panel) and AKI Stage 3 patients with CCI ≥ 3 and ICU stay ≥ 48 hours (right panel)

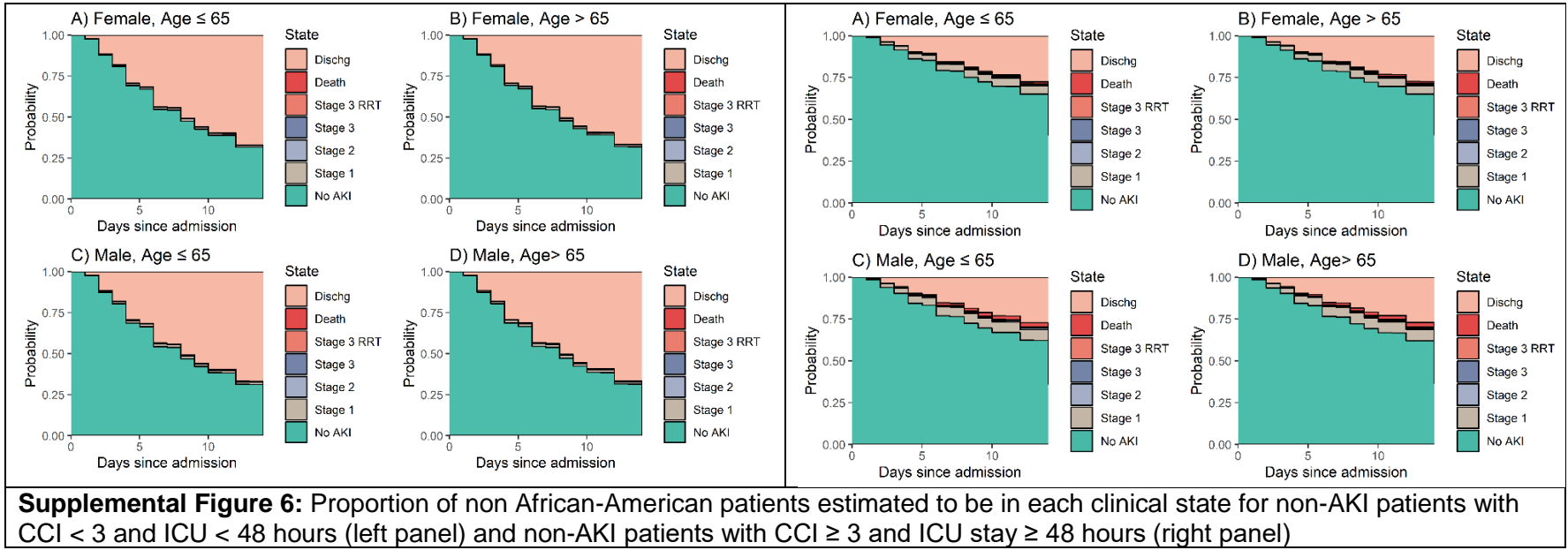

**Supplemental Figure 6:** Proportion of non African-American patients estimated to be in each clinical state for non-AKI patients with CCI < 3 and ICU < 48 hours (left panel) and non-AKI patients with CCI ≥ 3 and ICU stay ≥ 48 hours (right panel)

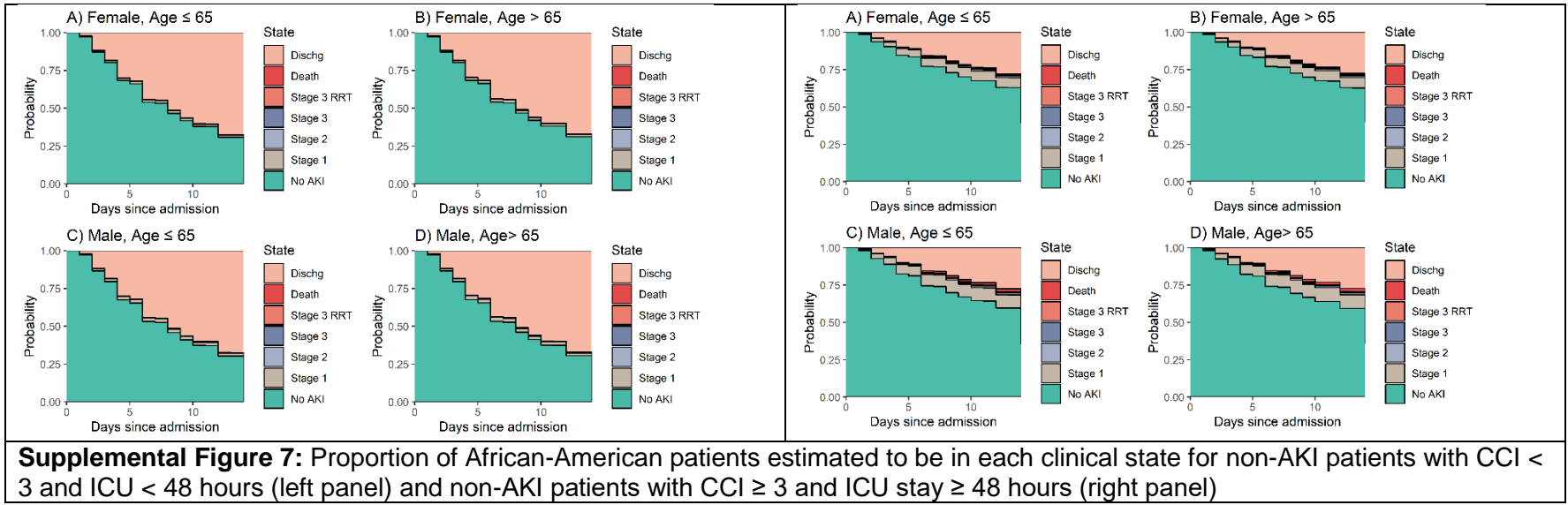

**Supplemental Figure 7:** Proportion of African-American patients estimated to be in each clinical state for non-AKI patients with CCI < 3 and ICU < 48 hours (left panel) and non-AKI patients with CCI ≥ 3 and ICU stay ≥ 48 hours (right panel)

**Supplemental Table 1. Regression coefficients for the model adjusted for age, sex, race, CCI ≥ 3 and ICU ≥ 48 hours (standard errors are given in parenthesis)**

| Transition | Age | Sex | Race | CCI≥3 | ICU≥48 hours | Transition | Age | Sex | Race | CCI≥3 | ICU≥48 hours |
|---|---|---|---|---|---|---|---|---|---|---|---|
| Admission to No AKI | -0.02 (0.00) | 0.00 (0.00) | -0.01 (0.01) | -0.05 (0.00) | -0.07 (0.01) | Stage 2 to Stage 1 | 0.11 (0.04) | -0.06 (0.04) | 0.09 (0.05) | -0.09 (0.04) | -0.58 (0.04) |
| Admission to AKI Stage 1 | 0.27 (0.02) | -0.09 (0.02) | 0.10 (0.02) | 0.40 (0.02) | 0.49 (0.02) | Stage 2 to Stage 3 | 0.06 (0.10) | 0.24 (0.10) | 0.22 (0.12) | 0.18 (0.10) | 0.09 (0.10) |
| Admission to AKI Stage 2 | 0.15 (0.03) | -0.11 (0.03) | 0.11 (0.04) | 0.53 (0.03) | 0.80 (0.03) | Stage 2 to Stage 3 + RRT | -0.46 (0.30) | 0.58 (0.29) | -0.90 (0.47) | -0.08 (0.28) | 1.16 (0.33) |
| Admission to AKI Stage 3 | -0.04 (0.03) | 0.14 (0.03) | 0.38 (0.04) | 0.72 (0.03) | 1.03 (0.03) | Stage 2 to Death | 0.45 (0.10) | 0.06 (0.10) | -0.15 (0.13) | 0.50 (0.11) | -0.05 (0.10) |
| Admission to AKI Stage 3 + RRT | -1.06 (0.28) | 0.44 (0.24) | -0.72 (0.36) | 1.17 (0.24) | 2.45 (0.28) | Stage 2 to Discharge | 0.00 (0.05) | -0.03 (0.05) | 0.04 (0.06) | -0.27 (0.05) | -1.02 (0.05) |
| No AKI to Stage 1 | 0.34 (0.02) | 0.07 (0.02) | 0.38 (0.02) | 0.60 (0.02) | 0.12 (0.02) | Stage 3 to No AKI | -0.07 (0.08) | 0.08 (0.08) | -0.06 (0.09) | -0.32 (0.08) | -0.78 (0.09) |
| No AKI to Stage 2 | -0.01 (0.07) | -0.10 (0.08) | 0.23 (0.08) | 0.43 (0.07) | 0.65 (0.07) | Stage 3 to Stage 1 | -0.03 (0.09) | -0.21 (0.09) | -0.17 (0.11) | -0.14 (0.10) | -0.59 (0.10) |
| No AKI to Stage 3 | 0.11 (0.12) | 0.17 (0.11) | 0.53 (0.13) | 0.43 (0.12) | 0.48 (0.12) | Stage 3 to Stage 2 | 0.17 (0.06) | -0.22 (0.06) | -0.24 (0.07) | -0.20 (0.06) | -0.48 (0.07) |
| No AKI to Stage 3 + RRT | 0.24 (0.27) | 0.62 (0.29) | -0.12 (0.37) | 0.84 (0.28) | 1.90 (0.31) | Stage 3 to Stage 3 + RRT | 0.04 (0.19) | 0.35 (0.19) | -0.39 (0.23) | -0.05 (0.18) | 1.03 (0.22) |
| No AKI to Death | 0.89 (0.04) | -0.01 (0.04) | -0.33 (0.06) | 0.63 (0.04) | 1.07 (0.04) | Stage 3 to Death | 0.34 (0.10) | 0.09 (0.10) | -0.13 (0.12) | 0.49 (0.11) | -0.19 (0.10) |
| No AKI to Discharge | 0.00 (0.00) | -0.01 (0.00) | 0.01 (0.01) | -0.26 (0.01) | -0.91 (0.01) | Stage 3 to Discharge | 0.03 (0.06) | 0.03 (0.05) | 0.10 (0.06) | -0.12 (0.05) | -0.94 (0.06) |
| Stage 1 to No AKI | 0.01 (0.02) | 0.04 (0.02) | 0.07 (0.02) | -0.10 (0.02) | -0.44 (0.02) | Stage 3 + RRT to No AKI | -6.70 (85.41) | -8.58 (72.84) | 11.04 (207.10) | -7.44 (71.51) | 7.04 (101.30) |
| Stage 1 to Stage 2 | 0.16 (0.06) | 0.02 (0.06) | 0.08 (0.07) | 0.09 (0.06) | 0.21 (0.06) | Stage 3 + RRT to Stage 1 | NA (0.00) | NA (0.00) | NA (0.00) | NA (0.00) | NA (0.00) |
| Stage 1 to Stage 3 | -0.16 (0.14) | -0.06 (0.13) | 0.23 (0.15) | 0.45 (0.14) | 0.67 (0.14) | Stage 3 + RRT to Stage 2 | NA (0.00) | NA (0.00) | NA (0.00) | NA (0.00) | NA (0.00) |

| Transition | Age | Sex | Race | CCI≥3 | ICU≥48 hours | Transition | Age | Sex | Race | CCI≥3 | ICU≥48 hours |
|---|---|---|---|---|---|---|---|---|---|---|---|
| **Stage 1 to Stage 3 + RRT** | -0.14 (0.28) | 0.40 (0.28) | -0.46 (0.39) | 0.04 (0.28) | **1.58** (0.32) | **Stage 3 + RRT to Stage 3** | **8.90** (67.58) | **8.15** (69.18) | **-5.60** (82.36) | **7.65** (61.15) | **7.30** (135.90) |
| **Stage 1 to Death** | **0.60** (0.08) | **0.24** (0.08) | **-0.33** (0.11) | **0.37** (0.08) | **0.45** (0.08) | **Stage 3 + RRT to Death** | 0.12 (0.15) | -0.08 (0.15) | 0.07 (0.20) | 0.14 (0.15) | **-0.87** (0.17) |
| **Stage 1 to Discharge** | **0.03** (0.02) | -0.02 (0.02) | **0.04** (0.02) | **-0.23** (0.02) | **-0.95** (0.02) | **Stage 3 + RRT to Discharge** | 0.06 (0.22) | -0.04 (0.22) | -0.47 (0.34) | -0.19 (0.21) | **-1.31** (0.24) |
| **Stage 2 to No AKI** | **-0.39** (0.05) | 0.05 (0.05) | 0.07 (0.06) | **-0.49** (0.05) | **-0.56** (0.05) | | | | | | |

Abbreviations. AKI, acute kidney injury; CCI, Charlson comorbidity index, ICU, intensive care unit.
Covariates with significance ≤ 0.05 are shown in bold face.